\documentclass[aps,pra,twocolumn,superscriptaddress,floatfix,nobibnotes]{revtex4-2}

\usepackage[utf8]{inputenc}
\usepackage[T1]{fontenc}
\usepackage{amsmath,amssymb,bm}
\usepackage{graphicx}
\usepackage{xcolor}
\usepackage{hyperref}
\usepackage{siunitx}
\usepackage{braket}
\usepackage{caption}
\usepackage{subcaption}
\usepackage{listings}
\usepackage{tikz}

\hypersetup{
  colorlinks=true,
  linkcolor=blue,
  citecolor=blue,
  urlcolor=blue
}
\newcommand{\re}[1]{\mathrm{Re}\left(#1\right)}
\newcommand{\im}[1]{\mathrm{Im}\left(#1\right)}

\newcommand{\brac}[1]{\left[ #1 \right]}
\newcommand{\pwisein}{\left\{ \begin{array}{ll}}
\newcommand{\pwiseout}{\end{array}\right.}

\newcommand{\trace}[1]{\mathrm{Tr} \left( #1 \right)}
\renewcommand{\det}[1]{\mathrm{det}\left( #1 \right)}

\begin{document}

\title{From triangles to prisms: towards a geometric extension of concurrence fill for three-qubit mixed states}
 
\author{Divyanshu K. Verma}
\affiliation{Department of Computer Science and Engineering, Indian Institute of Technology Ropar, Rupnagar - 140001, Punjab, India}
\author{Girish Kulkarni}
\affiliation{Department of Physics, Indian Institute of Technology Ropar, Rupnagar - 140001, Punjab, India}

\date{\today}

\begin{abstract}
The quantification of multipartite entanglement remains a long-standing open challenge in quantum information theory with major implications for quantum foundations and quantum technologies. In the context of three-qubit pure states, the concurrence triangle underlies geometric interpretations for a variety of multipartite entanglement measures. In particular, the concurrence fill, which is proportional to the square root of the area of the concurrence triangle, has been shown to be a valid measure of genuine tripartite entanglement. As a result, there is now significant interest in extending the concurrence triangle and concurrence fill to three-qubit mixed states. In this work, we show that any rank-2 three-qubit mixed state can be visualized in terms of concurrence prisms associated with their pure state decompositions. Thus, the concurrence fill of the mixed state obtained through convex-roof extension is proportional to the total prism fill area, defined as the square root of the base area times the height, associated with the prisms minimized over all pure state decompositions. We then consider three-qubit mixed states that are incoherent mixtures of Greenberger-Horne-Zeilinger (GHZ) and W states and analytically perform their convex-roof extension to obtain a closed-form expression for the concurrence fill of these rank-2 mixtures. Finally, we apply the geometric picture of concurrence prisms to these mixtures and their pure state decomposition corresponding to minimum concurrence fill as a tool for visualization.
\end{abstract}

\maketitle

\section{Introduction}

Multipartite entanglement, which refers to the inseparability of three or more quantum systems, exhibits several rich and complex features that are absent in its bipartite counterpart \cite{horodecki2024arxiv}. For instance, tripartite systems exhibit features such as entanglement monogamy \cite{coffman2000pra}, two inequivalent classes of entanglement \cite{dur2000pra}, and violation of local realism even at the level of single measurements \cite{Greenberger1989springer,greenberger1990AJP}, which have no analogs in bipartite systems. In addition, tripartite entanglement offers more security for teleportation compared to bipartite systems \cite{bennet1993prl, bennett1996prl,joo2003newjp,lee2005pra}. In view of such distinctive features, there has been significant interest in developing a rigorous quantitative understanding of multipartite entanglement and its implications for quantum foundations and applications \cite{shang2024review}. 

The quantification of bipartite entanglement is conceptually straightforward because every bipartite pure state has a unique Schmidt decomposition and every entanglement measure is some function of the Schmidt spectrum \cite{wootters1998prl,wootters2001qic,horodecki2009revmodphy}. In other words, all bipartite entanglement measures lead to the same ordering of states with respect to their degrees of entanglement because there is a single class of maximally entangled states to which all states are connected via local operations and classical communication (LOCC) \cite{nielsen1999prl,nielsen2000quantum}. In contrast, the quantification of multipartite entanglement is highly complicated. In general, there is no analog of the Schmidt decomposition for multipartite states indicating that there may not be a single class of entanglement measures \cite{horodecki2009revmodphy}. This is because there are multiple distinct inequivalent classes of maximally entangled states in multipartite systems that are not connected via stochastic LOCC \cite{dur2000pra,horodecki2009revmodphy}. For instance, for three-qubit systems, there are two distinct maximally entangled states, namely, the Greenberger-Horne-Zeilinger (GHZ) state and the W state \cite{dur2000pra,eibl2004prl,joo2003newjp}. In addition, one has to distinguish between fully-separable states, states that have only bipartite entanglement, namely, biseparable states, and states such as GHZ-type and W-type states that have genuine tripartite entanglement \cite{acin2001prl,dur2000pra,ma2011pra,das2016pra}. This distinction is important from a practical standpoint because three-party teleportation requires genuine tripartite entanglement as a resource \cite{boschi1998aps, bennett1996prl,joo2003newjp,lee2005pra}. As a result, efforts are underway to formulate a measure of genuine tripartite entanglement that satisfies two conditions: (i) it must be zero for fully-separable and biseparable states, and (ii) it must be non-zero for GHZ-type and W-type genuinely entangled states \cite{ma2011pra,das2016pra,sen2010pra}. 

At the heart of such efforts is a geometric object known as the concurrence triangle that is constructed with the three side lengths to be the squares of the concurrences for the three distinct possible bipartitions of a given three-qubit pure state \cite{xie2021prl}. Previously, several three-qubit entanglement measures were formulated, but none of them satisfied both conditions (i) and (ii) \cite{barnum2001jpa,ma2011pra,das2016pra,sen2010pra}. But recently, Xie and Eberly used the concurrence triangle to construct a measure known as the concurrence fill, which is proportional to the square root of the area of the concurrence triangle, that satisfies both conditions (i) and (ii) \cite{xie2021prl}. Moreover, the measure is monotonic under LOCC and ranks the W state to be less entangled than the GHZ state. The latter property is consistent with the fact that quantum teleportation using the W state is associated with lower success probability and average fidelity than that using the GHZ state \cite{joo2003newjp,lee2005pra}. In view of the utility of the concurrence triangle for visualization and the excellent properties of concurrence fill as a genuine entanglement measure, it would be useful to extend both of these concepts to three-qubit mixed states. 

In this paper, we extend the formulations of concurrence triangle and concurrence fill to three-qubit mixed states of rank two. In particular, we show that every pure state decomposition of a rank-2 three-qubit mixed state can be visualized in terms of a set of concurrence prisms associated with the decomposition. Moreover, the concurrence fill for the mixed state computed via convex roof extension\cite{osterloh2025arxiv,androulakis2022arxiv,uhlmann2010entropy,aggarwal2025classification} simply corresponds to a minimization of the total prism fill area -- which is defined for a given prism as the product of the height and the square root of the base area -- over all pure state decompositions.  We then consider rank-2 mixtures of GHZ and W states, and analytically perform their convex-roof extension to obtain a closed-form expression for the concurrence fill of these rank-2 mixtures. We show that the geometric picture of concurrence prisms to these mixtures provides a neat visualization tool and a geometric interpretation to the convex roof optimization procedure. 

The remainder of the paper is organized as follows: In Sec.~\ref{concurrenceprisms}, we introduce the concept of concurrence prisms and establish some notations. In Sec.~\ref{geometric}, we develop a geometrical picture that illustrates the connection between the eigenprisms corresponding to the eigendecomposition and the general prisms corresponding to the general pure state decomposition of a rank-2 three-qubit mixed state. In Sec.~\ref{ghzw}, we consider mixtures of GHZ and W states and analytically derive their concurrence fill. Here, we illustrate the application of the geometrical picture of concurrence prisms to such mixtures. In Sec.~\ref{conclusion}, we conclude with a summary and outlook of our study. 

\section{Concurrence Prisms}\label{concurrenceprisms}

We recall that for any three-qubit pure state $\ket{\psi_{ABC}}$, the concurrence triangle has the three side lengths equal to the squared concurrences $C^{2}_{A(BC)},C^{2}_{B(CA)},$ and $C^{2}_{C(AB)}$ corresponding to the three possible bipartitions. The concurrence fill $F_{ABC}$ of the state is proportional to the square root of the triangle area as \cite{xie2021prl}
\begin{align}\notag
F_{ABC} &= \left[ \frac{16}{3}Q\left(Q-C_{A(BC)}^2\right)\left(Q-C_{B(AC)}^2\right)\right.\\&\hspace{18mm}\left.\left(Q-C_{C(AB)}^2\right)\right]^{\frac{1}{4}}\label{concurrence fill},
\end{align}
where
\begin{equation}
Q = \frac{C_{A(BC)}^2+C_{B(AC)}^2+C_{C(AB)}^2}{2}
\end{equation}
is the semi-perimeter of the triangle. Also, the concurrence $C_{A(BC)}$ between qubit $A$ and the pair of qubits $BC$ can be evaluated as
\begin{equation}\label{concurrence}
C_{A(BC)} = 2 \sqrt{\det {\rho_A}},
\end{equation}
where $\rho_{A}$ is the reduced density matrix of qubit $A$ \cite{coffman2000pra}.
\begin{figure}[t]
    \centering    \includegraphics[width=0.65\columnwidth, angle=270]{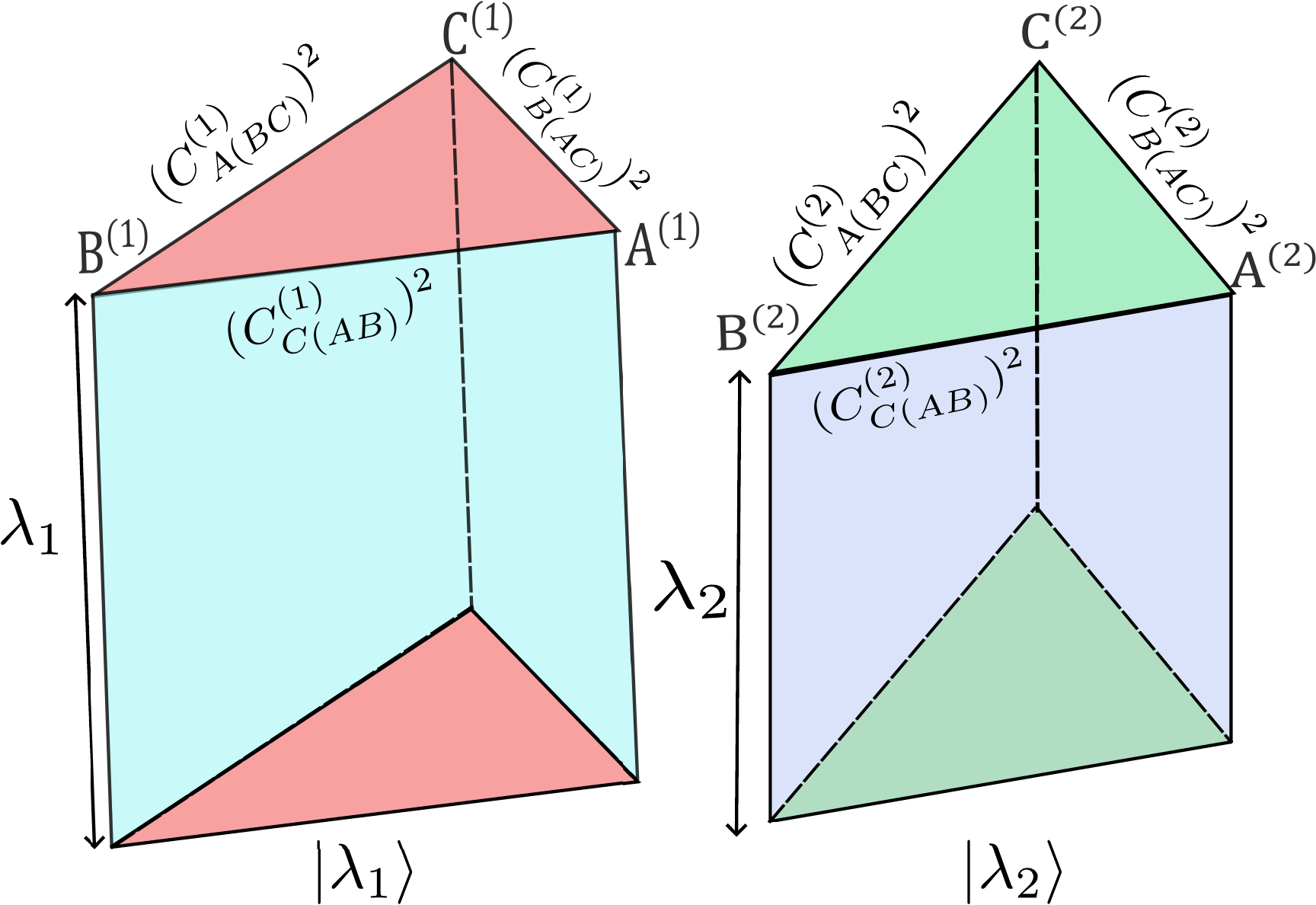}
    \caption{Eigenprisms: Concurrence prisms associated with the eigendecomposition.}
    \label{fig1}
\end{figure}

We now consider a rank-2 mixed state $\rho(\lambda)$ of three qubits $A,B$ and $C$ that has the eigendecomposition
\begin{equation}\label{eigendecomp}
\rho(\lambda) = \lambda_1\ket{\lambda_1}\bra{\lambda_1} + \lambda_2\ket{\lambda_2}\bra{\lambda_2},
\end{equation}
where $\lambda_1 = \lambda$ and $\lambda_2 = 1-\lambda$. For this decomposition, one can associate a set of concurrence prisms as shown in Fig.~\ref{fig1}, which we refer to as eigenprisms. Each prism has the concurrence triangle of the corresponding eigenstate as its base and the corresponding eigenvalue as its height. 

According to Caratheodory's theorem \cite{cook1972cmb}, the same state $\rho(\lambda)$ has a general pure state decomposition of the form
\begin{equation}
\rho(\lambda) = \sum_{k=1}^4 p_k \ket{\phi_k}\bra{\phi_k}.\label{gendecomp}
\end{equation}
Similarly, one can associate a set of four concurrence prisms, which we refer to as general prisms, to this general decomposition where the $k$'th prism has concurrence triangle for $|\phi_{k}\rangle$ as the base and $p_{k}$ as the height. Using convex roof extension, the concurrence fill of $\rho(\lambda)$ is given by
\begin{equation}\label{convexroof}
F_{ABC}\left(\rho(\lambda)\right) = \min \left\{\sum_{k=1}^4 p_k F_{ABC}\left(\ket{\phi_k}\right)\right\},
\end{equation}
where the minimization is carried out over all possible pure state decompositions.\\

\textbf{Definition of prism fill area:} We define the prism fill area $F$ of a prism as $F=h\sqrt{A}$, where $A$ is the base area and $h$ is the height of the prism. \\

We notice that the minimization in Eq. (\ref{convexroof}) essentially amounts to finding the pure state decomposition of $\rho(\lambda)$ whose concurrence prisms have the minimum total prism fill area. In other words, the convex roof minimization acquires an elegant geometric interpretation. 

In order to perform the minimization, it is useful to establish the connection between the eigendecomposition Eq. (\ref{eigendecomp}) and the general decomposition Eq. (\ref{gendecomp}). In this regard, we define a complex co-isometry $V$ as 
\begin{equation}\label{V}
V = \begin{pmatrix}
    v_{11} & v_{12} & v_{13} & v_{14}\\v_{21}&v_{22}&v_{23}&v_{24}
\end{pmatrix},
\end{equation}
using which one can obtain the following relations \cite{hughston1993pla},
\begin{subequations}
\begin{align}
p_k &= \lambda |{v_{1k}}|^2 + (1-\lambda)|v_{2k}|^2,\label{eq8a}\\
\ket{\phi_k} &= \frac{v_{1k}\sqrt{\lambda}\ket{\lambda_1}+v_{2k}\sqrt{1-\lambda}\ket{\lambda_2}}{\sqrt{p_k}}
.\label{eq8b}
\end{align}
\end{subequations}
In other words, every general pure state decomposition is uniquely related to the eigendecomposition through a co-isometry $V$ \cite{hughston1993pla}. Thus, the complex matrix elements $v_{ij}$ can be used as the free parameters over which the convex roof optimization of Eq. (\ref{convexroof}) can be carried out.  

We note that the state $|\phi_{k}\rangle$ in Eq. (\ref{eq8b}) can be written as
\begin{equation}\label{superposition}
 |\phi_{k}\rangle = \alpha_{k}\ket{\lambda_1}+\beta_{k}\ket{\lambda_2},
 \end{equation}
for $k=1,2,3,4$, where $\alpha_{k}=v_{1k}\sqrt{\lambda}/\sqrt{p_{k}}$ and $\beta_{k}=v_{2k}\sqrt{1-\lambda}/\sqrt{p_{k}}$ are complex coefficients that satisfy $|\alpha_{k}|^2+|\beta_{k}|^2=1$. In what follows, we compute the concurrence triangle sides of $|\phi_{k}\rangle$ in terms of the concurrence triangle sides of $\ket{\lambda_1}$ and $\ket{\lambda_2}$. To that end, we will first establish some notations. 
\begin{figure*}[t!]
	\centering
	\includegraphics[scale=0.55]{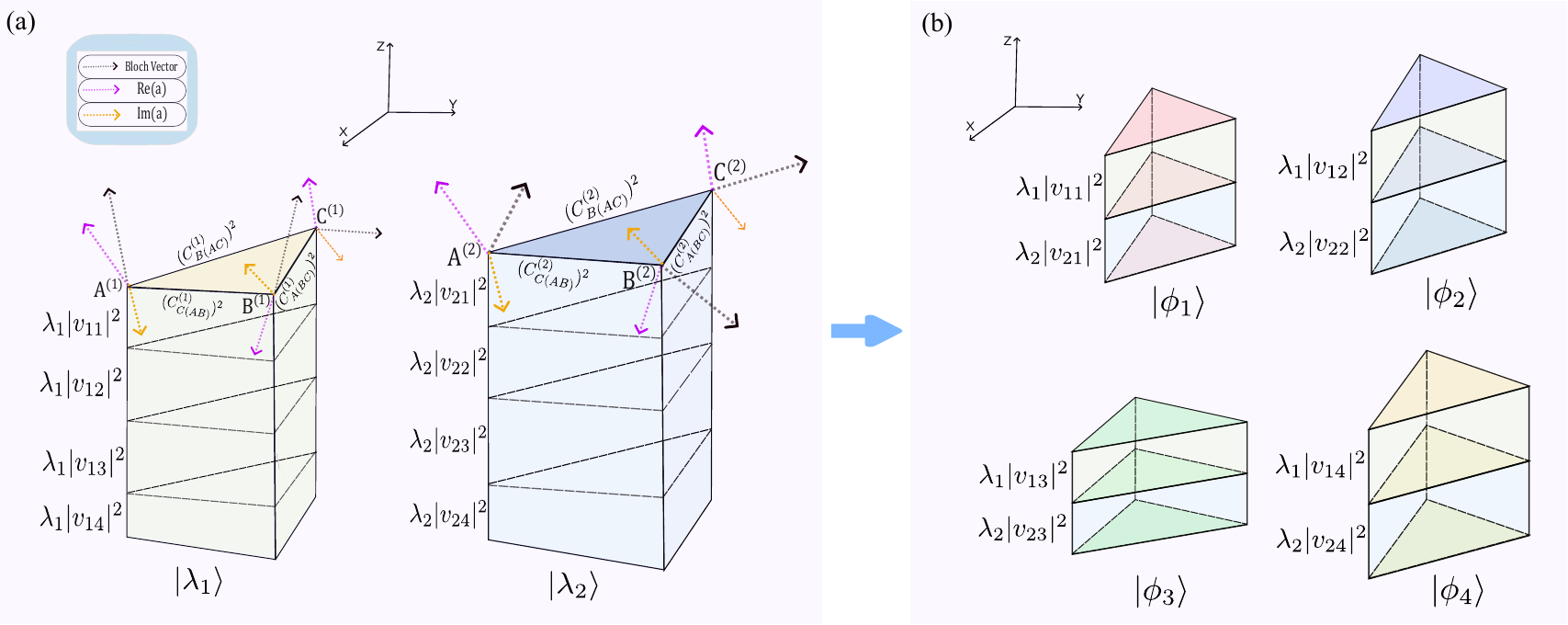}
	\caption{Schematic depiction of the geometrical procedure connecting the eigenprisms to the general prisms.}
	\label{demo}
\end{figure*}

We note that the eigenstate $|\lambda_{i}\rangle$ for $i=1,2$ can be represented by a column vector as
\begin{equation}
\ket{\lambda_{i}}=\begin{pmatrix}
    \mu^{(i)}_{000}\\\mu^{(i)}_{001}\\\mu^{(i)}_{010}\\\mu^{(i)}_{011}\\\mu^{(i)}_{100}\\\mu^{(i)}_{101}\\\mu^{(i)}_{110}\\\mu^{(i)}_{111}
\end{pmatrix},
\end{equation}
We now define the matrices
\begin{subequations}
\begin{align}
M^{(i)}_A&=\begin{pmatrix}
    \mu^{(i)}_{000}&\mu^{(i)}_{001}&\mu^{(i)}_{010}&\mu^{(i)}_{011}\\\mu^{(i)}_{100}&\mu^{(i)}_{101}&\mu^{(i)}_{110}&\mu^{(i)}_{111}
\end{pmatrix},\\
M^{(i)}_B&=\begin{pmatrix}
    \mu^{(i)}_{000}&\mu^{(i)}_{001}&\mu^{(i)}_{100}&\mu^{(i)}_{101}\\\mu^{(i)}_{010}&\mu^{(i)}_{011}&\mu^{(i)}_{110}&\mu^{(i)}_{111}
\end{pmatrix},\\
M^{(i)}_C&=\begin{pmatrix}
    \mu^{(i)}_{000}&\mu^{(i)}_{010}&\mu^{(i)}_{100}&\mu^{(i)}_{110}\\\mu^{(i)}_{001}&\mu^{(i)}_{011}&\mu^{(i)}_{101}&\mu^{(i)}_{111}
\end{pmatrix}.
\end{align}
\end{subequations}
In terms of these definitions, the reduced density matrices of the three qubits in the eigenstate $|\lambda_{i}\rangle$ take the form
\begin{equation} \label{eigen_reduced}
\Lambda^{(i)}_{j}=M^{(i)}_{j}M^{(i)\dagger}_{j},
\end{equation}
for $i=1,2$ and $j=A,B,$ and $C$. In other words, the matrices $M^{(i)}_{j}$ can be used to directly compute the reduced density matrices of the subsystems in the two eigenstates. Similarly, it is useful to define
\begin{equation}
X^{(1,2)}_{j} = \mathrm{Tr}_{kl}\ket{\lambda_1}\bra{\lambda_2}= M_{j}^{(1)}M_{j}^{(2)\dagger},
\end{equation}
where $(j,k,l)$ is a cyclic permutation of $(A,B,C)$. 

Using these notations, we compute the reduced density matrix $\rho^{(k)}_{j}$ of the $j$ subsystem in the global state $|\phi_{k}\rangle$ Eq.~\eqref{superposition} as
\begin{align}\notag
\rho^{(k)}_j &= |\alpha_{k}|^2\Lambda_j^{(1)} + |\beta_{k}|^2\Lambda_j^{(2)} +\alpha_{k}\beta_{k}^* X^{(1,2)}_j\\ &\hspace{38mm}+ \alpha_{k}^*\beta_{k} X_j^{(1,2)\dagger},
\end{align}
for $j=A,B,$ and $C$. Using the relation (\ref{concurrence}), it follows that $C^2_{l(mn)}\left(\ket{\phi_{k}}\right)=4\det{\rho^{(k)}_{l}}$, which upon simplification yields
\begin{align}\notag
&C_{l(mn)}^2\left(\ket{\phi_{k}}\right)= |\alpha_{k}|^4\left(C_{l(mn)}^{(1)}\right)^2 + |\beta_{k}|^4\left(C_{l(mn)}^{(2)}\right)^2 \\\notag
&\hspace{10mm} + 4|\alpha_{k}|^2|\beta_{k}|^2\left[1-\trace{\Lambda_l^{(1)}\Lambda_l^{(2)}}\right] \\\notag
&\hspace{10mm} - 4|\alpha_{k}\beta_{k}|^2\trace{X^{(1,2)}_lX_l^{\dagger(1,2)}} \\\notag
&\hspace{10mm} - 4\re{(\alpha_{k}\beta_{k}^*)^2\trace{X_l^{2(1,2)}}} \\\notag
&\hspace{10mm} - 8\re{|\alpha_{k}|^2\alpha_{k}\beta_{k}^*\trace{\Lambda_l^{(1)}X^{(1,2)}_l}} \\
&\hspace{10mm} - 8\re{|\beta_{k}|^2\alpha_{k}\beta_{k}^*\trace{\Lambda_l^{(2)}X^{(1,2)}_l}}
\label{side_conc}
\end{align}
for $l(mn)=A(BC),B(CA),$ and $C(AB)$. Thus, one can obtain the concurrence triangle sides of $|\phi_{k}\rangle$ in terms of the concurrence triangle sides of the eigenstates. We will now describe a geometric procedure that allows one to migrate from the concurrence prisms of the eigendecomposition Eq.~(\ref{eigendecomp}) to the general concurrence prisms of any general decomposition Eq.~(\ref{gendecomp}). In this procedure each term in the above equation (\ref{side_conc}) acquires an interesting geometrical interpretation.   

\section{From the eigenprisms to the general prisms: a geometric picture}\label{geometric}

We refer to Fig.~\ref{demo} to visualize the following steps in the procedure for constructing the general prisms.

We place the two eigenprisms with a non-zero arbitrary lateral distance between them and the axes of both being parallel to each other. We then set the direction along the heights of the prisms as the z-axis. For this z-axis, we arbitrarily assign orthonormal x and y axes in the plane perpendicular to it. We then employ this coordinate system independently to all the vertices of the eigenprisms with the origins of the resulting twelve systems at the twelve vertices. Each vertex thus gets an associated coordinate system.  

We then select six vertices, three each on the top triangular face, and label them as $j^{(i)}$, where $i=1,2$, referring to the first and second eigenprism, and $j=A,B,$ and $C$. We plot the Bloch vector $\mathbf{v}_j^{(i)}$ for the state $\Lambda^{(i)}_j$ of Eq.~(\ref{eigen_reduced}) on the coordinate system assigned to vertex $j^{(i)}$. Let us define $\mathbf{v}^{(i)}_{j}\in\mathbb{R}^3$ as the Bloch vector corresponding to the state $\Lambda_j^{(i)}$ \cite{nielsen2000quantum}, i.e,
\begin{equation}
    \Lambda_j^{(i)}=\frac{1}{2}\left(\mathbb{I}_2+\mathbf{v}^{(i)}_{j}\cdot\mathbf{\sigma}\right),
\end{equation}
where $\mathbf{\sigma}\equiv(\sigma_{x},\sigma_{y},\sigma_{z})$. One can then show that 
\begin{equation}
    \trace{\Lambda_j^{(1)}\Lambda_j^{(2)}}=\frac{1+\mathbf{v}_j^{(1)}\cdot\mathbf{v}_j^{(2)}}{2}.
    \label{eq17}
\end{equation}
Similarly, each matrix $X_j^{(1,2)}$ can be decomposed in the Pauli basis as
\begin{equation}
    X_j^{(1,2)}=\mathbf{a}_j\cdot\mathbf{\sigma}=\sum_{\mu=1}^3 a_{j\mu}\sigma_\mu,
\end{equation}
where $\mathbf{a}_j\in \mathbb{C}^3$ and $a_{j\mu}= \frac{1}{2}\trace{ X_j^{(1,2)}\sigma_\mu}$. We plot two copies of $\re{\mathbf{a}_j}$ and $\im{\mathbf{a}_j}$, one each on the coordinate systems associated with vertices $j^{(1)}$ and $j^{(2)}$ . 

Using the above definitions, one can obtain the following relations 
\begin{subequations}
\begin{align}
    \re{\trace{\Lambda_j^{(i)}X^{(1,2)}_j}} =\re{\mathbf{a}_j}\cdot\mathbf{v}_j^{(i)}, \\
    \im{\trace{\Lambda_j^{(i)}X^{(1,2)}_j}}=\im{\mathbf{a}_j}\cdot\mathbf{v}_j^{(i)}.
\end{align}
\begin{equation}
    \trace{X^{(1,2)}_jX^{^\dagger(1,2)}_j}=2|\mathbf{a}_j|^2.
    \label{eq22}
\end{equation}
\begin{equation}
    \trace{(X^{(1,2)}_j)^2}=2(\mathbf{a}_j\cdot\mathbf{a}_j)=2\sum_{\mu=0}^3a_{j\mu}^2,
\end{equation}
\begin{align}\notag
    \trace{(X^{(1,2)}_j)^2}=2(|\re{\mathbf{a}}|^2-|\im{\mathbf{a}}&|^2)&\\+4i(\re{\mathbf{a}}\cdot\im{\mathbf{a}}).
\end{align}
\end{subequations}
This construction accompanied by the co-isometry $V$ provides a geometrical method to construct concurrence triangles for pure states of Eq.~\eqref{eq8b} by providing a geometrical meaning to each term in Eq.~\eqref{side_conc}. To extend the four triangles obtained into general prisms, we divide each eigenprism into four horizontal sections. The height of the $k^{th}$ section will be $|v_{1k}|^2\lambda_1$ and $|v_{2k}|^2\lambda_2$ for eigenprisms 1 and 2 respectively, where $v_{mn}$ are the elements of the co-isometry $V$ defined in Eq.~(\ref{V}). For the $k^{th}$ general prism, the height will be obtained by adding the heights of the $k^{th}$ horizontal section of both eigenprisms, $|v_{1k}|^2\lambda_1+|v_{2k}|^2\lambda_2$. So following Eq.~\eqref{eq8a}, the height of each general prism corresponds to the probability weight of the corresponding pure state in the pure state decomposition. Thus, the total prism fill area of the general prisms $p_1F_{ABC}(\ket{\phi_1})+p_2F_{ABC}(\ket{\phi_2})+p_3F_{ABC}(\ket{\phi_3})+p_4F_{ABC}(\ket{\phi_4})$ is essentially proportional to the average concurrence fill of the pure state decomposition.

\section{Application to mixtures of GHZ and W states}\label{ghzw}

\begin{figure}[t!]
\centering
\includegraphics[width=1\columnwidth]{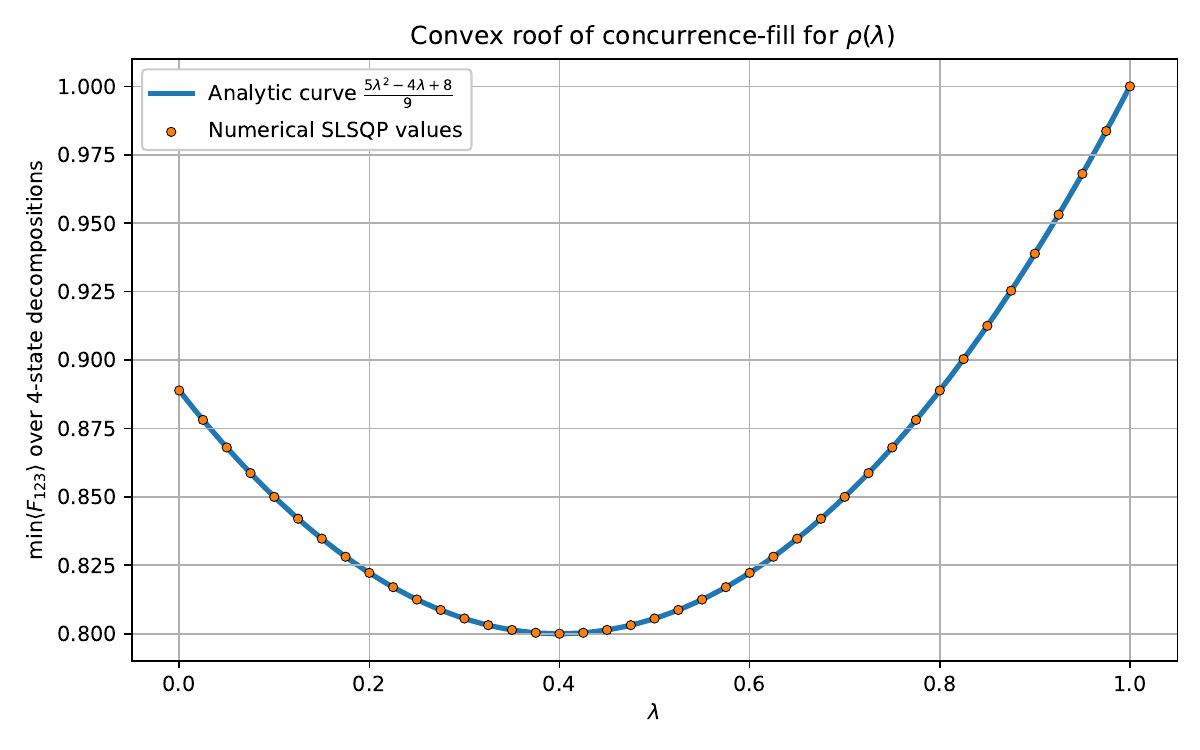}
\caption{Concurrence fill of rank-2 mixtures of the form $\rho(\lambda) = \lambda\ket{GHZ}\bra{GHZ} + (1-\lambda )\ket{W}\bra{W}$ with respect to the parameter $\lambda$.}
\label{graph}
\end{figure}

We consider a mixed state $\rho(\lambda)$ of the form
\begin{equation}
    \rho(\lambda) = \lambda\ket{GHZ}\bra{GHZ} + (1-\lambda )\ket{W}\bra{W}, \label{GHZ-W state}
\end{equation}
where $\lambda$ is a real parameter such that $0\leq \lambda\leq 1$, $\ket{GHZ} = \left(\ket{000}+\ket{111}\right)/\sqrt{2}$ and $\ket{W} = \left(\ket{001}+\ket{010}+\ket{100}\right)/\sqrt{3}$ are the orthonormal eigenstates.
Using our established notations, any pure-state decomposition of $\rho(\lambda)$ expressed by Eq.~\eqref{gendecomp} and defined by a co-isometry $V$ of Eq.~\eqref{V} is obtained by using Eqs.~\eqref{eq8a} and \eqref{eq8b} as
\begin{subequations}
\begin{align}
p_k &= \lambda |{v_{1k}}|^2 + (1-\lambda)|v_{2k}|^2,\label{eq27a}\\
\ket{\phi_k} &= \frac{v_{1k}\sqrt{\lambda}\ket{GHZ}+v_{2k}\sqrt{1-\lambda}\ket{W}}{\sqrt{p_k}}
.\label{eq27b}
\end{align}
\end{subequations}
For any such decomposition, the average concurrence fill is $F_{ABC}^{avg}(\rho(\lambda))=\sum_{k=1}^4 p_kF_{ABC}(\ket{\phi_k})$. 

Since Eq.~\eqref{eq27b} is essentially a coherent superposition of the GHZ and W eigenstates, let us consider a general pure state of the form 
\begin{equation} \label{GHZ-W_pure}
\ket{\psi}=\sqrt{q}\ket{GHZ}+\sqrt{1-q}\ket{W},
\end{equation}
where $q \in [0,1]$. For this three-qubit pure state $\ket{\psi}$, the reduced density matrices of the subsystems take the form
\begin{equation}  
\rho_j =\begin{pmatrix}
    \frac{4-q}{6}&-\sqrt{\frac{q(1-q)}{6}}\\\sqrt{\frac{q(1-q)}{6}}&\frac{2+q}{6}
\end{pmatrix},
\end{equation}
which implies that the concurrence triangle is equilateral with each side having length
\begin{align}
\begin{split}
    C_{l(mn)}^2 = 4\det{\rho_j} =\frac{5q^2 - 4q +8}{9}.
\end{split}
\end{align}
Thus, the concurrence fill of $\ket{\psi}$ according to Eq.~\eqref{concurrence fill} is
\begin{align}\notag
    F_{ABC}(\ket{\phi}) &= \brac{\frac{16}{3} \cdot\frac{3}{2}C_{l(mn)}^2\cdot\frac{1}{8}C_{l(mn)}^6}^\frac{1}{4} = C_{l(mn)}^2\\&=\frac{5q^2 - 4q +8}{9}.
\end{align}
Comparing Eq.~\eqref{GHZ-W_pure} with Eq.~\eqref{eq27b}, we set $q = \frac{|v_{1k}^2|\lambda}{{p_k}}$ to get 
\begin{equation}
    F_{ABC}(\ket{\phi_k}) = \frac{5\lambda^2{|v_{1k}|^4}}{9p_k^2} - \frac{4\lambda|v_{1k}|^2}{9p_k}+\frac{8}{9}.
\end{equation}
Consequently, the average concurrence fill takes the form,
\begin{align}\notag
    {F_{ABC}^{avg}(\rho(\lambda))} &= \sum_{k=1}^4 p_k\left[\frac{5\lambda^2{|v_{1k}|^4}}{9p_k^2}\right.\\&\hspace{20mm}\left.- \frac{4\lambda|v_{1k}|^2}{9p_k}+\frac{8}{9}\right].
\end{align}
We now need to find the minimum of $F_{ABC}^{avg}(\rho)$ over all possible pure state decompositions to quantify the concurrence fill of $\rho(\lambda)$. To that end, let us denote $r_k = \frac{|v_{1k}|^2}{p_k}$ such that
\begin{equation}
    {F_{123}^{avg}(\rho)} = \sum_{k=1}^4 p_k\brac{\frac{5\lambda^2r_k^2}{9} - \frac{4\lambda r_k}{9}+\frac{8}{9}}.
\end{equation}
Let us define a function $g(r_k)$ as
\begin{equation}
    g(r_k) =\frac {5\lambda^2r_k^2}{9} - \frac{4\lambda r_k}{9}+\frac{8}{9}.
\end{equation}
In other words, we are trying to minimize the function
\begin{equation} \label{F}
    F = \sum_k p_kg(r_k) = \mathbb{E}_p[g(r_k)],
\end{equation}
where $\mathbb{E}_p[g(r_k)]$ denotes the expectation value of $g(r_k)$. Also note that $p_kr_k = |v_{1k}|^2$ and since $V$ is a co-isometry,
\begin{equation}
    \sum_kp_kr_k = \mathbb{E}_p[r_k] = \sum_k|v_{1k}|^2 = 1.
\end{equation}
As $g"(r_k)>0$, it follows that $g(r_k)$ is a convex function of $r_k$. According to Jensen's inequality for convex functions \cite{mcshane1937bams}
\begin{equation}
    \mathbb{E}_p[g(r_k)] \geq g(\mathbb{E}_p[r_k]) = g(1).
\end{equation}
Consequently, from Eq.~\eqref{F} it follows that,
\begin{equation}
    F \geq g(1) =  \frac{5\lambda^2}{9} - \frac{4\lambda }{9}+\frac{8}{9},
\end{equation}
In other words, if we can find a pure state decomposition, or equivalently, a co-isometry $V$ such that $F$ achieves the minimum value of $g(1)$, then one can say that it is one of the pure state decompositions achieving the minimum average concurrence fill for $\rho(\lambda)$. 

In fact, many such co-isometries $V$ are possible, one of which is
\begin{equation}
    V = \begin{pmatrix}
    1/2 & 1/2 & 1/2 & 1/2\\1/2&1/2&-1/2&-1/2
\end{pmatrix},
\label{eq40}
\end{equation}
from which it can be concluded that
\begin{equation}
    \min F_{ABC}(\rho(\lambda)) = \frac{5\lambda^2}{9} - \frac{4\lambda }{9}+\frac{8}{9}.
    \label{41}
\end{equation}
This result is verified by performing the approximate convex-roof construction for a range of values of $\lambda$ and finding the concurrence fill for each corresponding mixed state in Eq.~\eqref{GHZ-W state} using numerical simulations, described in Fig.~\ref{graph}. The dotted points show the minimal average $\langle F_{ABC}\rangle$ found by constrained Sequential Least Squares Programming (SLSQP) minimisation over ensembles as a function of the mixing parameter $\lambda$ (uniform grid of 41 points).  Each point was obtained with a multi-start procedure (40 random initialisations) using SLSQP (\texttt{ftol}=1e-9, \texttt{maxiter}=1500); constraint residuals were checked to ensure feasibility.  The bold plot depicts the curve $F_{\min}(\lambda)=(5\lambda^2-4\lambda+8)/9$ and perfectly overlaps with the numerical points. Thus, our theoretical result of Eq.~\eqref{41} is perfectly consistent with numerical simulations. 
\begin{figure*}[t!]
    \centering    \includegraphics[width=0.95\linewidth]{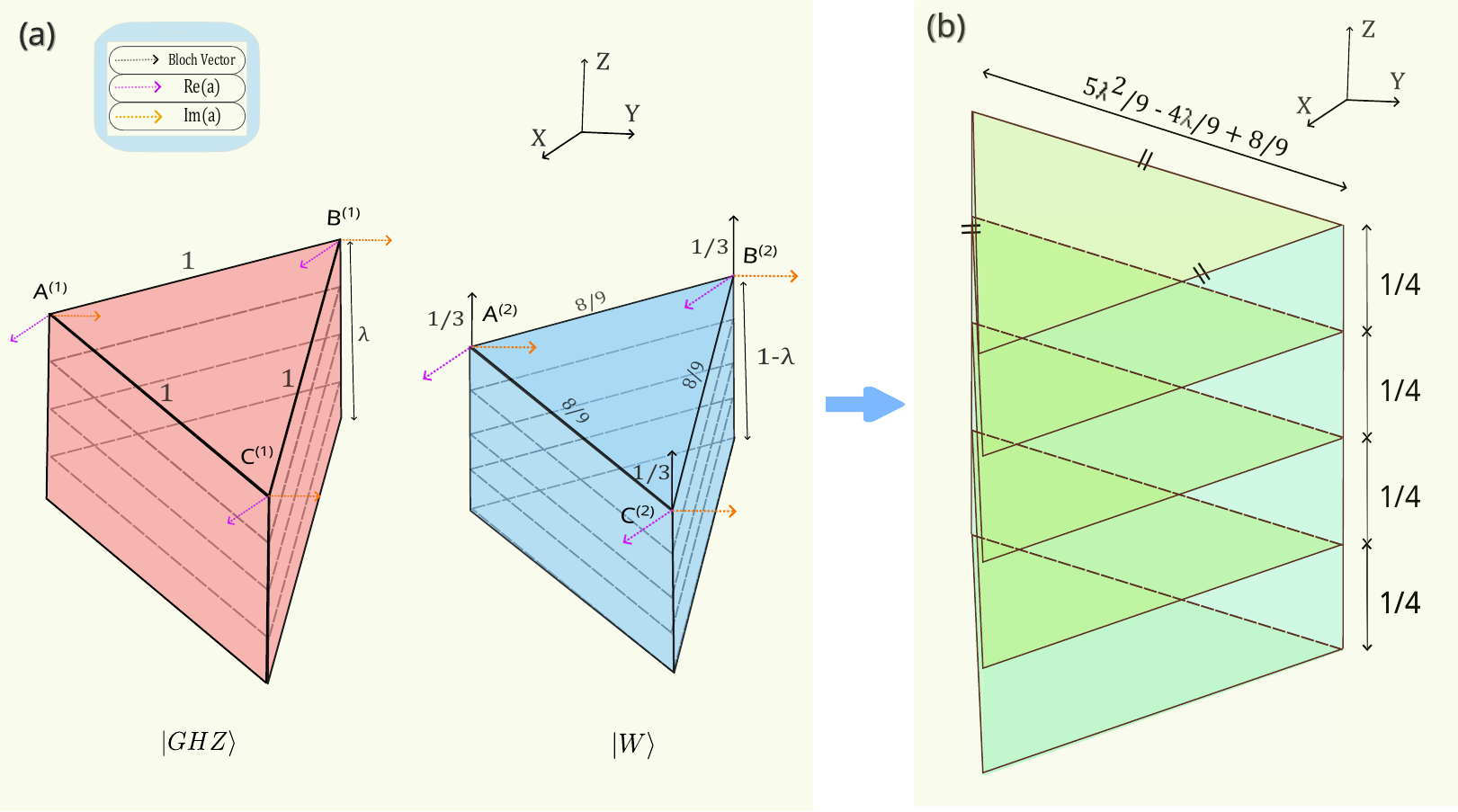}
    \caption{$(a)$ The eigenprisms for the GHZ-W mixture with heights $\lambda$ and $(1-\lambda)$ respectively. Following from the co-isometry $V$ minimizing the total prism fill area of the general prisms, each transverse section is divided equally. $(b)$ The identical general prisms obtained following the procedure described in Sec.~\ref{geometric}.}
    \label{ghzwfig}
\end{figure*}

The concurrence prism picture for $\rho(\lambda)$ is constructed in Fig.~\ref{ghzwfig} by evaluating the following quantities (see Appendix \hyperref[appendix_a]{A})
\begin{subequations}
\begin{align}\label{calc}
        &C^2_{l(mn)}(\ket{GHZ})=1\\&C^2_{l(mn)}(\ket{W})=\frac{8}{9}\\&\mathbf{v}_j^{GHZ}=(0,0,0)\\&
\mathbf{v}_j^{W}=(0,0,1/3)\\&
\mathrm{Tr}(\Lambda_j^{GHZ}\Lambda_j^{W})=\frac{1}{2}\\&
\mathrm{Tr}(X_j^{(GHZ,W)}X_j^{\dagger(GHZ,W)}) = \frac{1}{6}\\&
\mathrm{Tr}(X_j^{2(GHZ,W)})=0\\&
\mathrm{Tr}(\Lambda_
j^{GHZ}X_j^{(GHZ,W)})=0\\&
\mathrm{Tr}(\Lambda_
j^{W}X_j^{(GHZ,W)})=0
\end{align} 
\end{subequations}
The co-isometry $V$ found in Eq.~\eqref{eq40} divides the eigenprisms into 4 equal horizontal sections, and the resulting general prisms are all identical Fig.~\ref{ghzwfig}. The total prism fill area of these general prisms correspond to the minimum concurrence fill evaluated previously in Eq.~\eqref{41}, depending only on the heights of the eigenprisms, which in turn is governed by the parameter $\lambda$. 

This example illustrates how the concurrence prism picture can be used as a visualization tool, thereby mapping the abstract algebraic optimization problem to an intuitive geometric optimization problem. 
\section{Summary and Outlook}\label{conclusion}

We extend the formulations of concurrence triangle and concurrence fill to three-qubit mixed states of rank two. In particular, we show that every pure state decomposition of a rank-2 three-qubit mixed state can be visualized in terms of a set of concurrence prisms associated with the decomposition. Moreover, the concurrence fill of the mixed state computed via convex roof extension simply corresponds to a minimization of the total prism fill area -- which is defined for a given prism as the product of the square root of the base area and the height -- over all pure state decompositions. In order to perform this optimization, it is useful to define a co-isometry that connects the eigendecomposition to the general pure state decomposition and carry out the optimization over the co-isometry parameters. We then consider rank-2 mixtures of GHZ and W states, and analytically perform their convex-roof extension to obtain a closed-form expression for the concurrence fill of these rank-2 mixtures. We show that the geometric picture of concurrence prisms provides a neat visualization tool and a geometric interpretation to the convex roof optimization procedure. 

We expect our work to be applicable to experiments on multipartite entanglement such as Ref.~\cite{farias2012prl, wang2016sb, mivcuda2017sr}. In the future, one could explore if the geometric picture of concurrence prisms and the minimization of prism fill area can be used to develop geometric algorithms that can find the optimal pure state decompositions for arbitrary rank-2 three-qubit mixed states \cite{regula2018jpa,vidal2002prl}. Furthermore, one could explore the extension of the same construction to general mixed states of higher rank. We hope that our work can lead to further insights into multipartite entanglement.

\section*{Acknowledgments}
We are grateful to Manik Banik for several insightful discussions during this work. We acknowledge financial support through the National Quantum Mission (NQM) of the Department of
Science and Technology, Government of India and the
initiation research grant received from IIT Ropar.

\appendix

\section{Explicit Calculations for the GHZ-W Mixture}

\label{appendix_a}

The target three-qubit states are defined as:
\begin{align*}
|GHZ\rangle &= \frac{1}{\sqrt{2}}(|000\rangle + |111\rangle), \\
|W\rangle &= \frac{1}{\sqrt{3}}(|001\rangle + |010\rangle + |100\rangle).
\end{align*}
Due to the permutation symmetry of these states, operations with respect to the subsystem $A$ is shown, the rest are implied.

For the GHZ state:
\begin{align*}
\rho_A^{GHZ} &= \text{Tr}_{BC}(|GHZ\rangle\langle GHZ|) \\
&= \text{Tr}_{BC}\bigg (\frac{1}{2} \Big( \ket{000}\bra{000} + \ket{000}\bra{111} \\
&\qquad\qquad\qquad\quad + \ket{111}\bra{000} + \ket{111}\bra{111} \Big) \bigg) \\
&= \frac{1}{2}(\ket{0}\bra{0} + \ket{1}\bra{1}) \\
&= \begin{pmatrix} \frac{1}{2} & 0 \\ 0 & \frac{1}{2} \end{pmatrix} \\
&= \frac{1}{2}\mathbb{I}_2.
\end{align*}
For the W state:
\begin{align*}
\rho_A^{W} &= \text{Tr}_{BC}(\ket{W}\bra{W}) \\ 
&= \text{Tr}_{BC}\bigg( \frac{1}{3}\Big( \ket{001}\bra{001} + \ket{001}\bra{010} + \ket{001}\bra{100} \\ 
&\qquad\qquad\quad + \ket{010}\bra{001} + \ket{010}\bra{010} + \ket{010}\bra{100} \\ 
&\qquad\qquad\quad + \ket{100}\bra{001} + \ket{100}\bra{010} + \ket{100}\bra{100} \Big) \bigg) \\ 
&=\frac{2}{3}|0\rangle\langle 0| + \frac{1}{3}|1\rangle\langle 1| \\ 
&= \begin{pmatrix} \frac{2}{3} & 0 \\ 0 & \frac{1}{3} \end{pmatrix}
\end{align*}
The squared concurrence across the bipartition $A(BC)$ is given by $C_{A(BC)}^2 = 4\det{\rho_A}$
\begin{align*}
C_{A(BC)}^2(\ket{GHZ}) &= 4 \text{det}{\begin{pmatrix} \frac{1}{2} & 0 \\ 0 & \frac{1}{2} \end{pmatrix}} = 4 \left(\frac{1}{4}\right) = 1 \\
C_{A(BC)}^2(\ket{W}) &= 4 \text{det}{\begin{pmatrix} \frac{2}{3} & 0 \\ 0 & \frac{1}{3} \end{pmatrix}} = 4 \left(\frac{2}{9}\right) = \frac{8}{9}
\end{align*}
Any single-qubit density matrix can be expanded in terms of its Bloch vector $\vec{v}$ as $\rho = \frac{1}{2}(\mathbb{I}_2 + \vec{v}\cdot\vec{\sigma})$.
For $\rho_A^{GHZ} = \frac{1}{2}\mathbb{I}_2$, the Bloch vector is trivially a null vector:
\begin{equation*}
\bar{v}_1 = (0, 0, 0). 
\end{equation*}
For $\rho_A^W$, we equate components:
\begin{equation*}
\frac{1}{2} \begin{pmatrix} 1+v_z & v_x - iv_y \\ v_x + iv_y & 1-v_z \end{pmatrix} = \begin{pmatrix} \frac{2}{3} & 0 \\ 0 & \frac{1}{3} \end{pmatrix}.
\end{equation*}
This implies $v_x = 0$, $v_y = 0$, and $\frac{1}{2}(1+v_z) = \frac{2}{3} \implies v_z = \frac{1}{3}$,
\begin{equation*}
\bar{v}_2 = \left(0, 0, \frac{1}{3}\right).
\end{equation*}
Next we calculate the trace of the product of the two reduced density matrices:
\begin{equation*}
\begin{aligned}
\operatorname{Tr}(\rho_A^{GHZ}\rho_A^{W})
&= \operatorname{Tr}\!\left(
\begin{bmatrix}
\frac{1}{2} & 0 \\
0 & \frac{1}{2}
\end{bmatrix}
\begin{bmatrix}
\frac{2}{3} & 0 \\
0 & \frac{1}{3}
\end{bmatrix}
\right) \\
&= \frac{1}{2}\cdot\frac{2}{3}
 + \frac{1}{2}\cdot\frac{1}{3}
= \frac{1}{3}+\frac{1}{6}
= \frac{1}{2}.
\end{aligned}
\end{equation*}
The matrix $X_A = \text{Tr}_{BC}(|GHZ\rangle\langle W|)$,
\begin{align*}
|GHZ\rangle\langle W| &= \frac{1}{\sqrt{6}} (|000\rangle + |111\rangle)(\langle 001| + \langle 010| + \langle 100|),
\end{align*}
\begin{equation*}
X_A = \frac{1}{\sqrt{6}} |0\rangle\langle 1| \underbrace{\langle 00|00\rangle}_{=1} = \frac{1}{\sqrt{6}} |0\rangle\langle 1| = \begin{pmatrix} 0 & \frac{1}{\sqrt{6}} \\ 0 & 0 \end{pmatrix}.
\end{equation*}
We map $X_A$ to a complex vector in $\mathbb{C}^3$ using $X = \vec{a}\cdot\vec{\sigma}$.
\begin{equation*}
\begin{pmatrix} a_3 & a_1 - i a_2 \\ a_1 + i a_2 & a_3 \end{pmatrix} = \begin{pmatrix} 0 & \frac{1}{\sqrt{6}} \\ 0 & 0 \end{pmatrix}.
\end{equation*}
From the diagonal elements, $a_3 = 0$. From the off-diagonal elements:
\begin{equation*}
a_1 - i a_2 = \frac{1}{\sqrt{6}}, \quad a_1 + i a_2 = 0 \implies a_1 = -i a_2.
\end{equation*}
Substituting $a_1$ into the first relation gives $-2i a_2 = \frac{1}{\sqrt{6}} \implies a_2 = \frac{i}{2\sqrt{6}}$ and $a_1 = \frac{1}{2\sqrt{6}}$.
Thus, the complex vector $\vec{a}$ corresponding to $X$ is:
\begin{equation*}
X = \frac{1}{\sqrt{6}}|0\rangle\langle1| \quad \rightarrow \quad \vec{a} = \left( \frac{1}{2\sqrt{6}}, \frac{i}{2\sqrt{6}}, 0 \right). 
\end{equation*}
Now, evaluating the various traces involving $X_A$:
\begin{equation*}
\text{Tr}(X X^\dagger) = \text{Tr}\left[ \begin{pmatrix} 0 & \frac{1}{\sqrt{6}} \\ 0 & 0 \end{pmatrix} \begin{pmatrix} 0 & 0 \\ \frac{1}{\sqrt{6}} & 0 \end{pmatrix} \right] = \text{Tr}\begin{pmatrix} \frac{1}{6} & 0 \\ 0 & 0 \end{pmatrix} = \frac{1}{6}. 
\end{equation*}
\begin{equation*}
\text{Tr}(X^2) = \text{Tr}\left[ \begin{pmatrix} 0 & \frac{1}{\sqrt{6}} \\ 0 & 0 \end{pmatrix} \begin{pmatrix} 0 & \frac{1}{\sqrt{6}} \\ 0 & 0 \end{pmatrix} \right] = \text{Tr}\begin{pmatrix} 0 & 0 \\ 0 & 0 \end{pmatrix} = 0. 
\end{equation*}
For the overlap between the reduced states and $X_A$:
\begin{equation*}
\text{Tr}(\rho_A^{GHZ}X) = \text{Tr}\left[ \begin{pmatrix} \frac{1}{2} & 0 \\ 0 & \frac{1}{2} \end{pmatrix} \begin{pmatrix} 0 & \frac{1}{\sqrt{6}} \\ 0 & 0 \end{pmatrix} \right] = \text{Tr}\begin{pmatrix} 0 & \frac{1}{2\sqrt{6}} \\ 0 & 0 \end{pmatrix} = 0. 
\end{equation*}
\begin{equation*}
\text{Tr}(\rho_A^{W}X) = \text{Tr}\left[ \begin{pmatrix} \frac{2}{3} & 0 \\ 0 & \frac{1}{3} \end{pmatrix} \begin{pmatrix} 0 & \frac{1}{\sqrt{6}} \\ 0 & 0 \end{pmatrix} \right] = \text{Tr}\begin{pmatrix} 0 & \frac{2}{3\sqrt{6}} \\ 0 & 0 \end{pmatrix} = 0. 
\end{equation*}

\bibliographystyle{apsrev4-2}
\bibliography{draft_entanglement_refs_clean}
\end{document}